\documentclass{svproc}

\usepackage[table]{xcolor}
\usepackage{times}
\usepackage{url}
\usepackage{graphicx}
\usepackage{amsmath}

\begin{document}
\mainmatter              
\title{RoboCup 2D Soccer Simulation League: Evaluation Challenges}
\titlerunning{RoboCup 2D Soccer Simulation League: Evaluation Challenges}  
%
\author{Mikhail Prokopenko$^1$, Peter Wang$^{2}$, Sebastian Marian$^{3}$, Aijun Bai$^{4}$, Xiao Li$^{5}$ \and  Xiaoping Chen$^{5}$}
\authorrunning{Prokopenko et al.} 
\institute{Complex Systems Research Group, Faculty of Engineering and IT\\ 
			The University of Sydney, NSW 2006, Australia\\
\email{mikhail.prokopenko@sydney.edu.au}\\
\and
Data Mining, CSIRO Data61, PO Box 76, Epping, NSW 1710, Australia\\
\and
Compa-IT, Romania\\
\and
Department of Electrical Engineering and Computer Sciences\\ 
University of California Berkeley\\
\and
Multi-Agent Systems Lab, School of Computer Science and Technology,\\
University of Science and Technology of China}

\maketitle              

\begin{abstract}
We summarise the results of RoboCup 2D Soccer Simulation League in 2016 (Leipzig), including the main competition and the evaluation round. The evaluation round held in Leipzig confirmed the strength of RoboCup-2015 champion (WrightEagle, i.e. WE2015) in the League, with only eventual finalists of 2016 competition  capable of defeating WE2015.  An extended, post-Leipzig, round-robin tournament which included the top 8 teams of 2016, as well as WE2015, with over 1000 games played for each pair, placed WE2015 third behind the champion team (Gliders2016) and the runner-up (HELIOS2016). This establishes WE2015 as a stable benchmark for the 2D Simulation League.  We then contrast two ranking methods and suggest two options for future evaluation challenges. The first one, ``The Champions Simulation League'', is proposed to include 6 previous champions, directly competing against each other in a round-robin tournament, with the view to systematically trace the advancements in the League. The second proposal, ``The Global Challenge'', is aimed to increase the realism of the environmental conditions during the simulated games, by simulating specific features of different participating countries.  
\end{abstract}

\section{Introduction}

The International RoboCup Federation's Millennium challenge sets an inspirational target that by mid-21st century, a team of fully autonomous humanoid soccer players shall win the soccer game, complying with the official rule of the FIFA, against the winner of the most recent World Cup  \cite{Burkhard}. In pursuit of this goal, the RoboCup  Federation has introduced multiple leagues, with both physical robots and simulation agents, which have developed different measures of their progress over the years. The main mode, of course, is running competitions at the national, regional and world cup levels. In addition, however, various leagues have included specific evaluation challenges which not only complement the competitions, but also advance the scientific and technological base of RoboCup and Artificial Intelligence in general. Typically, a challenge introduces some new features into the standard competition environment, and then evaluates how the teams perform under the new circumstances.

For example, during an evaluation round of RoboCup 2001 the rules of the soccer simulator were modified in such a way that ``dashing on the upper half of the field resulted in only half of normal speed for all the players'' \cite{Obst04}. This modification was not announced in advance, and while the changed conditions were obvious to human spectators, none of the simulation agents could diagnose the problem  \cite{Obst04}. 

A specific technical challenge was presented by the so-called Keepaway problem \cite{LNAI2005-keepaway}, when one team (the ``keepers'') attempt to keep the ball away from the other team (the ``takers'') for as long as possible.

Later on, the focus of evaluation in RoboCup 2D Soccer Simulation League shifted from changing the physics of the simulation or the tactics of the game, to studying the diverse ``eco-system'' of the League itself, which has grown to include multiple teams. The Simulated Soccer Internet League (SSIL) was designed to allow a continual evaluation of the participating teams during the time between annual RoboCup events: pre-registered teams could upload their binaries to a server on which games were played automatically \cite{Obst02}. The SSIL was used at some stage as a qualification pathway to the annual RoboCup, but this practice was discontinued due to verification problems.

Several other challenges and technical innovations introduced in Soccer Simulation Leagues (both 2D and 3D), including heterogeneous players, stamina capacity model, and tackles, are described in \cite{AkiyamaDL14}. This study further pointed out the importance of the online game analysis and online adaptation.

More recently, a series of ``drop-in player challenges'' was introduced by \cite{MacAlpine2014} in order to investigate how real or simulated robots from teams from
around the world can cooperate with a variety of unknown teammates. In each evaluation game, robots/agents are drawn from the participating teams and combined to form a new team, in the hope that the agents would be able to quickly adapt to meaningfully play together without pre-coordination.  The ``drop-in'' challenge was adopted by RoboCup Standard Platform League (SPL) and both RoboCup Soccer Simulation Leagues, 2D and 3D.  In all the considered leagues, the study observed ``a trend for agents that perform better at standard team soccer to also perform better at the drop-in player challenge'' \cite{MacAlpine2014}.  

At RoboCup-2016 in Leipzig, several soccer and rescue leagues increased realism of the competition by holding their competitions outdoors. In the SPL, a separate competition was successfully held not on the customary green carpet  but rather on an artificial turf, under diverse natural lighting conditions. Similarly, Middle Size Soccer League  also successfully implemented a Technical Challenge under these difficult conditions, while the Humanoid League used artificial turf and real soccer balls\footnote{http://robocup2016.org/press-releases/leipzig-best-place-for-robots-and-friends/452749}.

In this paper, we describe the latest evaluation challenge, introduced by RoboCup 2D Soccer Simulation League \cite{Kitano97,Noda2003} in 2016, in order to trace the progress of the overall League. Furthermore, we describe two possibilities for future challenges: one intended to systematically trace the advancements in the League (``The Champions Simulation League''), and the other aimed to increase the realism of the environmental conditions during the simulated games (``The Global Challenge'').

\section{Methodology and Results}

\subsection{Actual competition}

The RoboCup-2016 Soccer Simulation 2D League  included 18 teams from 9 countries: Australia, Brazil, China, Egypt, Germany, Iran, Japan, Portugal and Romania. The last group stage was a round-robin tournament for top 8 teams. It was followed by the two-game semi-final round, a single-game final, and 3 more playoffs between third and fourth, fifth and sixth, and seventh and eighth places. 

In the two-game semi-final round, HELIOS2016 (Japan) \cite{helios16} defeated team Ri-one (Japan) \cite{rione16}, 3:0 and 4:0, while Gliders2016 (Australia) \cite{gliders2016tdp,cyb98gliders16} defeated team CSU\_Yunlu (China) \cite{yunlu16},  winning both games with the same score 2:1.  

The single-game final between  HELIOS2016 and Gliders2016 went into the extra time, and ended with Gliders2016 winning 2:1.  

The third place was taken by team Ri-one which won against CSU\_Yunlu 3:0. 

Oxsy (Romania) \cite{oxsy16} took the fifth place, winning 4:0 against Shiraz (Iran) \cite{shiraz16}; and MT2016 (China) \cite{mt16} became seventh, winning against FURY (Iran) \cite{fury16} on penalties 4:2.  The final ranking of RoboCup-2016 (Leipzig, Germany) is shown in the left column of Table \ref{t1}.

\subsection{Ranking Estimation}

Using the ranking estimation methodology established by \cite{rcsymp2014,RAM15}, we conducted an 8-team round-robin tournament for top 8 teams from RoboCup-2016. The estimation process used the released binaries of top RoboCup-2016 teams\footnote{https://chaosscripting.net/files/competitions/RoboCup/WorldCup/2016/2DSim/binaries/}, where all 28 pairs of teams play approximately 4000 games against one another. The following \emph{discrete} scheme was used for discrete point calculation:

\begin{itemize}
\item Firstly, the average score between each pair of teams (across all 4000 games) is rounded to the nearest integer (e.g. ``1.2 : 0.5" is rounded to ``1 : 1"). 
\item Next, points are allocated for each pairing based on these rounded results: 3 for a win, 1 for a draw and 0 for a loss. 
\item Teams are then ranked by the sum of the points received against each opponent. The total goal difference of the rounded scores is used as a tie-breaker.
\end{itemize}

\noindent The final ranking $\mathbf{r^d}$ under this scheme is presented in Table \ref{t1}. 

\begin{table}[h]
\begin{center}
\begin{tabular}{|c|c|c|c|c|c|c|c|c|c|c|c|c|}
 \hline
 & \scriptsize{Gliders} &  \scriptsize{HELIOS} & \scriptsize{Ri-one} & \scriptsize{CSU\_Yunlu} & \scriptsize{Oxsy} & \ \scriptsize{Shiraz} \ &  \scriptsize{MT2016}  & \scriptsize{FURY}  & \scriptsize{Goals} & \scriptsize{Points} & \scriptsize{$\mathbf{r^d}$} \\ \hline
\scriptsize{Gliders} & \cellcolor[gray]{0.4} & \scriptsize{0.3 : 0.4} & \scriptsize{2.8 : 0.3}  & \scriptsize{1.9 : 0.3}  & \scriptsize{0.7 : 0.8} & \scriptsize{3.8 : 0.4}  &  \scriptsize{5.0 : 0.0}  & \scriptsize{2.5 : 0.2}  & \scriptsize{18 : 1} & \scriptsize{17} & \scriptsize{1} \\ \hline
\scriptsize{HELIOS} & \scriptsize{0.4 : 0.3}  & \cellcolor[gray]{0.4} & \scriptsize{1.8 : 0.1}  & \scriptsize{3.0 : 0.2}  & \scriptsize{1.2 :	0.5}  & \scriptsize{4.3 : 0.3}  & \scriptsize{3.6 : 0.0}  & \scriptsize{2.5 : 0.0}   & \scriptsize{17 : 1} & \scriptsize{17} & \scriptsize{2}\\ \hline
\scriptsize{Ri-one} & \scriptsize{0.3 : 2.8} & \scriptsize{0.1 : 1.8} & \cellcolor[gray]{0.4} &  \scriptsize{1.1 : 1.1}  &  \scriptsize{0.2 : 1.8}  & \scriptsize{0.6 : 0.5}  & \scriptsize{0.4 : 0.0}  & \scriptsize{0.6 : 0.5}   & \scriptsize{3 : 10}  & \scriptsize{4} & \scriptsize{6}\\ \hline
\scriptsize{CSU\_Yunlu} & \scriptsize{0.3 : 1.9} & \scriptsize{0.2 : 3.0} &  \scriptsize{1.1 : 1.1}  & \cellcolor[gray]{0.4} & \scriptsize{0.5 : 1.2}  & 	\scriptsize{2.0 : 0.7}  & \scriptsize{1.4 : 0.0}  & \scriptsize{1.2 : 0.4}  & \scriptsize{6 : 8} & \scriptsize{11} & \scriptsize{4} \\ \hline
\scriptsize{Oxsy}  & \scriptsize{0.8 : 0.7} & \scriptsize{0.5 : 1.2} & \scriptsize{1.8 : 0.2}  & \scriptsize{1.2 : 0.5} & \cellcolor[gray]{0.4} & \scriptsize{3.5 : 0.5}  & 	\scriptsize{4.4 : 0.0}  & 	\scriptsize{3.0 : 0.1}  &  \scriptsize{16 : 4}  & \scriptsize{15} & \scriptsize{3} \\ \hline
\scriptsize{Shiraz} & \scriptsize{0.4 : 3.8} & \scriptsize{0.3 : 4.3} & \scriptsize{0.5 : 0.6} & \scriptsize{0.7 : 2.0} & \scriptsize{0.5 : 3.5} & \cellcolor[gray]{0.4} & \scriptsize{0.5 : 0.1} & \scriptsize{0.8 : 1.0} & \scriptsize{5 : 16}  & \scriptsize{5} & \scriptsize{5}\\ \hline
\scriptsize{MT2016} & \scriptsize{0.0 : 5.0} & \scriptsize{0.0 : 3.6} & \scriptsize{0.0 : 0.4} & \scriptsize{0.0 : 1.4}  & \scriptsize{0.0 : 4.4}  & \scriptsize{0.1 : 0.5}  & \cellcolor[gray]{0.4} & \scriptsize{0.0 : 0.0}   &   \scriptsize{0 : 15}  & \scriptsize{2} & \scriptsize{8} \\ \hline
 \scriptsize{FURY}  & \scriptsize{0.2 : 2.5} &  \scriptsize{0.0 : 2.5} & \scriptsize{0.5 : 0.6} & \scriptsize{0.4 : 1.2} & \scriptsize{0.1 : 3.0} & \scriptsize{1.0 : 0.8}  & \scriptsize{0.0 : 0.0} & \cellcolor[gray]{0.4}  & \scriptsize{2 : 12} & \scriptsize{3} & \scriptsize{7} \\ \hline
\end{tabular}
\end{center}
\caption{Round-robin results (average goals scored and points allocated with the discrete scheme) for the top 8 teams from RoboCup 2016, ordered according to their final actual competition rank, $\mathbf{r^a}$. The scores are determined by calculating the average number of goals scored over approximately 4000 games rounded to the nearest integer, then awarding 3 points for a win, 1 point for a draw and 0 points for a loss. The resultant ranking is marked with $\mathbf{r^d}$.}
\label{t1} \vspace*{-8mm}
\end{table}

In order to capture the overall difference between any two rankings $\mathbf{r^a}$ and $\mathbf{r^b}$, the $L_1$ distance is utilised \cite{rcsymp2014}:
\begin{equation}
\label{eq:1}
    d_1(\mathbf{r^a}, \mathbf{r^b}) = \|\mathbf{r^a} - \mathbf{r^b}\|_1 = \sum_{i=1}^n |r^a_i-r^b_i| \ ,
\end{equation}
\noindent{where $i$ is the index of the $i$-th team in each ranking, $1 \leq i \leq 8$. 

The distance between the actual ranking $\mathbf{r^a}$ and the estimated ranking $\mathbf{r^d}$ is 
\begin{equation*}
    d_1(\mathbf{r^{a}}, \mathbf{r^{d}}) = |1 - 1| + |2 - 2| + |3 - 6| + |4 - 4| + |5 - 3| + |6 - 5| + |7 - 8| + |8 - 7| =  8.
\end{equation*}

The top two teams were fairly close in their performance (confirmed by the final game, which needed extra time). Similarly the 7th and 8th teams were similar in strength too (not surprisingly their playoff ended up with penalties).  The main discrepancy between the actual and estimated  results is due to performances of two teams: Oxsy (whose rank is  estimated as third, while the actual rank was only fifth) and Ri-one (which finished the competition as third, while its average rank is estimated to be  sixth).  

\vspace*{-2mm}
\subsection{A critique of the continuous ranking scheme}

There exists another ranking method: \emph{continuous} scheme \cite{rcsymp2014,RAM15}:
\begin{itemize}
\item Teams are ranked by the sum of average points obtained against each opponent across all 4000 games.
\item The total goal difference of the non-rounded scores is used as a tie-breaker.
\end{itemize}

Both schemes, discrete and continuous, were introduced in order to evaluate different competition formats, using the top 8 teams of 2012 and 2013 \cite{rcsymp2014,RAM15}. However, over the years it has become apparent that the continuous scheme suffers from two major drawbacks, violating the balance of points (3 for a win, 1 for a draw and 0 for a loss) and overestimating the points for draws and losses. Specifically, under the  continuous scheme:
 
\begin{enumerate}
\item there is a bias to attribute more points to draws with higher scores.
\item there is a bias to reduce the advantage of the three-points-for-a-win standard.
\end{enumerate}

1. Let us consider two opposite scenarios: (i) two teams $A$ and $B$ of equal strengths, denoted $A \Leftrightarrow B$, but with a stronger defensive capability, play $N$ games resulting in the average $0:0$ score; and (ii) two teams $X$ and $Y$ of equal strengths $X \Leftrightarrow Y$, but with a stronger attacking capability,  play $N$ games resulting in the average $q:q$ score, where $q > 0 $ is sufficiently large, e.g., $q=3$.  In the first pair, the scores of individual games, which may or may not be draws, do not diverge much from $0:0$, as the teams are defensive. And so the actual drawn scores $0:0$ dominate among the results, with large outliers $k:0$ or $0:k$, for $k > 0$ being relatively rare. Thus, the continuous points $p$ attained by teams $A$ and $B$ stay close to $1.0$, for example, $p_A \approx p_B \approx  1.2$. 

In the second pair, the scores of individual games, which again may or may not be draws,  diverge more from the average $q:q$, due to a higher variability of possible high scores. Consequently, the proportion of actual draws among $N$ games is much smaller in comparison to the first pair, and the large outliers $k:0$ or $0:k$, even for $k > q$, are more numerous. As a result, the teams $X$ and $Y$ exchange wins and losses more often than teams $A$ and $B$, acquiring more points for their respective wins. This yields the continuous points $p_X$ and $p_Y$ significantly higher than $1.0$, for example, $p_X \approx p_Y \approx 1.4$, creating a general bias to  attribute more points to the drawn contests with higher scores: $p_A \approx p_B < p_X \approx p_Y$.  A typical sample of 10,000 scores $q_1:q_2$, where both $q_1$ and $q_2$ are normally distributed around the same mean $q$, with the standard deviation $\sigma = 1.0$, results in the following continuous points $p_{\Leftrightarrow}(q)$ for different draws around $q$: $p_{\Leftrightarrow}(0) = 1.23$ for draws $0.38 : 0.38$, $p_{\Leftrightarrow}(1) = 1.33$ for draws $1.07 : 1.08$, $p_{\Leftrightarrow}(2) = 1.36$ for draws $1.99 : 2.00$, and $p_{\Leftrightarrow}(3) = 1.38$ for draws $3.02 : 3.00$.

While the higher scoring teams may be expected to get an advantage at a tie-breaker stage, getting more continuous points for the same outcome is obviously an unfair bias.  The discrete scheme does not suffer from this drawback as the average scores are converted into the identical discrete points immediately, i.e., $p_A = p_B = p_X = p_Y = 1.0$.

It is easy to see that the lower bound for the continuous points shared by any two teams of equal strength is $\inf_{\Leftrightarrow}  = 1.0$ (attainable only if all $N$ games are drawn), while the upper bound is $\sup_{\Leftrightarrow} = 1.5$ (attained in the extreme case when all $N$ games are non-draws, with wins and losses split equally).  Consequently, under the continuous scheme, the points attributed to equal teams drawing on average, are overestimated, being somewhere between the lower and upper bounds: $\inf_{\Leftrightarrow} < p < \sup_{\Leftrightarrow}$, while the expected result (one point) sits only at exactly the lower bound.

2.  The ``three-points-for-a-win'' standard which was widely adopted since FIFA 1994 World Cup finals ``places additional value on wins with respect to draws such that teams with a higher number of wins may rank higher in tables than teams with a lower number of wins but more draws''\footnote{https://en.wikipedia.org/wiki/Three\_points\_for\_a\_win}.  To illustrate the second drawback of the continuous scheme we will contrast two scenarios, comparing the combined points of two drawn contests against the combination of one-won and one-lost contests.  

Firstly, we consider a case when team $Q$ is paired with teams $U$ and $Z$, such that $Q \Leftrightarrow U$ and $Q \Leftrightarrow Z$. We do not expect transitivity, and so $U \Leftrightarrow Z$ is not assumed. The  continuous points for team $Q$ resulting from these two iterated match-ups, both drawn, could vary between these lower bound ($\inf_{\Leftrightarrow,\Leftrightarrow}$) and  upper bound ($ \sup_{\Leftrightarrow,\Leftrightarrow}$):
\begin{gather*}
 \inf_{\Leftrightarrow,\Leftrightarrow}  =  \inf_{\Leftrightarrow} + \inf_{\Leftrightarrow} = 2.0  \\
 \sup_{\Leftrightarrow,\Leftrightarrow}  = \sup_{\Leftrightarrow} + \sup_{\Leftrightarrow} = 3.0  
\end{gather*}
Typically the combined points vary around the level of $p_Q \approx 2.6$, which is an overestimation of the ideal outcome by more than half-a-point. 

Secondly, we consider a scenario with team $R$ matched-up against teams $V$ and $W$, with team $V$ being weaker than $R$, denoted  $R \Rightarrow V$, while the team $W$ is stronger than $R$, denoted $R \Leftarrow W$.  The relative strength of $V$ and $W$ is not important for our comparison.  The continuous points that team $R$ attains from the first pair, against the weaker opponent $V$, are bounded by $\inf_{\Rightarrow}  = 1.5$ (just a slight over-performance) and $\sup_{\Rightarrow}  = 3.0$ (the total dominance with all $N$ games won):
\[ 1.5 = \inf_{\Rightarrow} < p_R < \sup_{\Rightarrow}  = 3.0 \ .\] 
In practice,  the stronger team rarely drops below $p_R \approx 2.0$ points.   In the second pair, team $R$ is weaker, and its continuous points are bounded by $\inf_{\Leftarrow}  = 0.0$ (the total inferiority with all $N$ games lost) and $\sup_{\Leftarrow}  = 1.5$ (getting almost to an equal standing):
\[ 0.0 = \inf_{\Leftarrow} < p_R < \sup_{\Leftarrow}  = 1.5 \ .\] 
In practice,  the weaker team rarely reaches beyond  $p_R \approx 1.0$ points. A typical sample of 10,000 scores $q_1:q_2$, where $q_1$ and $q_2$ are normally distributed around the means $q$ and $0.0$ respectively, with the standard deviation $\sigma = 1.0$, results in the following continuous points $p_{\Rightarrow}(q)$ for different won contests around $q$: $p_{\Rightarrow}(1) = 2.31$ for wins $1.07 : 0.38$, $p_{\Rightarrow}(2) = 2.75$ for wins $2.00 : 0.38$, and $p_{\Rightarrow}(3) = 2.94$ for wins $2.97 : 0.38$. Correspondingly, the continuous points $p(q)$ for the respective lost contests sampled under the same distribution are overestimated above $0.0$ as follows: $p_{\Leftarrow}(1) = 0.32$, $p_{\Leftarrow}(2) = 0.13$, and $p_{\Leftarrow}(3) = 0.04$.

The combined continuous points for team $R$ after these match-ups, one won and one lost, could vary between the lower bound of and the upper bound of 
\begin{gather*}
\inf_{\Rightarrow,\Leftarrow}  = \inf_{\Rightarrow} + \inf_{\Leftarrow} = 1.5 \\
\sup_{\Rightarrow,\Leftarrow}  = \sup_{\Rightarrow} + \sup_{\Leftarrow} = 4.5 
\end{gather*}
In practice, $2.0 < p_R < 4.0$. That is, the combined continuous points of a win and a loss typically vary around  $p_R \approx 3.0$, which is an appropriate outcome. 

Contrasting the possible bounded intervals and typical outcomes of two contests (two draws versus one win and one loss) immediately highlights that the continuous points do not differentiate these scenarios sufficiently well. The intention of the three-points-for-a-win standard was precisely to preference the one-win-and-one-loss scenario over the two-draws scenario, $p_{\Rightarrow,\Leftarrow} = 3 > p_{\Leftrightarrow,\Leftrightarrow} = 2$. In other words, team $Q$ with two drawn contests should achieve a lower rank than team $R$ with a won and a lost contest, with the difference being the cost of a single drawn game. The continuous scheme fails in this regard, by producing, on average, less than half-a-point difference, $p_{\Rightarrow,\Leftarrow} \approx 3.0 > p_{\Leftrightarrow,\Leftrightarrow} \approx 2.6$.  
In fact, it is quite conceivable that $p_{\Rightarrow,\Leftarrow}$ could happen to be less than $p_{\Leftrightarrow,\Leftrightarrow}$ under the continuous scheme in some cases, as $\inf_{\Rightarrow,\Leftarrow} < \sup_{\Leftrightarrow,\Leftrightarrow}$. In other words, one hard-won contest, e.g. $p_{\Rightarrow}(1) = 2.31$, coupled with a serious loss, e.g., $p_{\Leftarrow}(3) = 0.04$ could earn less points (e.g., $p_{\Rightarrow,\Leftarrow} \approx 2.35$) than two high-scoring draws, e.g. $p_{\Leftrightarrow}(3) = 1.38$ (resulting in $p_{\Leftrightarrow,\Leftrightarrow} \approx 2.76$) --- definitely, something not intended by the three-points-for-a-win standard.

Again, the discrete scheme easily overcomes this drawback as the average scores are converted into the appropriate discrete points for each contest (3 for a win, 1 for a draw and 0 for a loss), and combined only afterwards. 

The two problems identified for the continuous scheme may amplify over many match-ups in a 8-teams round-robin, especially when there are many teams of similar strength (which is the case in the Simulation League in recent years). The biases become even more pronounced in the absence of transitivity in teams' relative strengths. In light of these concerns, we suggest that  some recent works  employing the continuous scheme, e.g. \cite{gabel2016}, would benefit from re-evaluation.

\subsection{Evaluation round}

The 2016 competition also included an evaluation round, where all 18 participating teams played one game each against the champion of RoboCup-2015, team WrightEagle (China), i.e., WE2015  \cite{we15}.  Only two teams, the eventual finalists Gliders2016 and HELIOS2016, managed to win against the previous year champion, with Gliders defeating WrightEagle 1:0, and HELIOS2016 producing the top score 2:1. 

We extended this evaluation over 1000 games, again playing  WE2015  against the top 8 teams from RoboCup-2016.  Table \ref{te} summarises the evaluation for RoboCup-2016: both actual scores obtained in Leipzig and the averages over 1000 games.

\vspace*{-3mm}
\begin{table}[h]
\begin{center}
\begin{tabular}{|c|c|c|c|c|c|c|c|c|c|}
 \hline
 & \scriptsize{Gliders2016} &  \scriptsize{HELIOS2016} & \scriptsize{Ri-one} & \scriptsize{CSU\_Yunlu} & \scriptsize{Oxsy} & \ \scriptsize{Shiraz2016} \ &  \scriptsize{MT2016}  & \scriptsize{FURY}   \\ \hline
\scriptsize{WE2015} & \scriptsize{0 : 1} & \scriptsize{1 : 2} & \scriptsize{7 : 1}  & \scriptsize{2 : 0}  & \scriptsize{4 : 1} & \scriptsize{3 : 2}  &  \scriptsize{4 : 0}  & \scriptsize{11 : 2}   \\ \hline
\scriptsize{WE2015} & \scriptsize{1.4 : 1.8} & \scriptsize{1.3 : 1.7} & \scriptsize{5.0 : 0.5}  & \scriptsize{2.7 : 0.5}  & \scriptsize{3.5 : 1.3} & \scriptsize{4.0 : 0.8}  &  \scriptsize{5.9 : 0.0}  & \scriptsize{4.8 : 0.4} \\
 \hline
\end{tabular}
\end{center}
\caption{Evaluation round results for the top 8 teams playing against WE2015. Top row: actual scores obtained at RoboCup-2016 in Leipzig; bottom row: average scores over 1000 games.}
\label{te}
\end{table}

\vspace*{-7mm}
The evaluation round confirmed the strength of RoboCup-2015 champion in the League.  It is evident that WE2015, if entered in 2016, would likely have achieved third rank.  To confirm this conjecture we combined the estimation results presented in Table \ref{t1} with the estimates of WE2015 scores from Table \ref{te}, summarised in Table \ref{te1}.

\vspace*{-3mm}
\begin{table}[h]
\begin{center}
\begin{tabular}{|c|c|c|c|c|c|c|c|c|c|c|c|c|c|}
 \hline
 & \scriptsize{Gliders} &  \scriptsize{HELIOS} & \scriptsize{WE2015} & \scriptsize{Ri-one} & \scriptsize{CSU\_Yunlu} & \scriptsize{Oxsy} & \ \scriptsize{Shiraz} \ &  \scriptsize{MT2016}  & \scriptsize{FURY}  & \scriptsize{Goals} & \scriptsize{Points} & \scriptsize{$\mathbf{r^{e}}$} \\ \hline
\scriptsize{Gliders} & \cellcolor[gray]{0.4} & \scriptsize{0.3 : 0.4} & \scriptsize{1.8 : 1.4}  & \scriptsize{2.8 : 0.3}  & \scriptsize{1.9 : 0.3}  & \scriptsize{0.7 : 0.8} & \scriptsize{3.8 : 0.4}  &  \scriptsize{5.0 : 0.0}  & \scriptsize{2.5 : 0.2}  & \scriptsize{20 : 2} & \scriptsize{20} & \scriptsize{1} \\ \hline
\scriptsize{HELIOS} & \scriptsize{0.4 : 0.3}  & \cellcolor[gray]{0.4} & \scriptsize{1.7 : 1.3} & \scriptsize{1.8 : 0.1}  & \scriptsize{3.0 : 0.2}  & \scriptsize{1.2 :	0.5}  & \scriptsize{4.3 : 0.3}  & \scriptsize{3.6 : 0.0}  & \scriptsize{2.5 : 0.0}   & \scriptsize{19 : 2} & \scriptsize{20} & \scriptsize{2}\\ \hline
\scriptsize{WE2015} & \scriptsize{1.4 : 1.8}  & \scriptsize{1.3 : 1.7} & \cellcolor[gray]{0.4}  & \scriptsize{5.0 : 0.5}  & \scriptsize{2.7 : 0.5}  & \scriptsize{3.5 :	1.3}  & \scriptsize{4.0 : 0.8}  & \scriptsize{5.9 : 0.0}  & \scriptsize{4.8 : 0.4}   & \scriptsize{29 : 8} & \scriptsize{18} & \scriptsize{3}\\ \hline
\scriptsize{Ri-one} & \scriptsize{0.3 : 2.8} & \scriptsize{0.1 : 1.8} & \scriptsize{0.5 : 5.0} & \cellcolor[gray]{0.4} &  \scriptsize{1.1 : 1.1}  &  \scriptsize{0.2 : 1.8}  & \scriptsize{0.6 : 0.5}  & \scriptsize{0.4 : 0.0}  & \scriptsize{0.6 : 0.5}   & \scriptsize{4 : 15}  & \scriptsize{4} & \scriptsize{7}\\ \hline
\scriptsize{CSU\_Yunlu} & \scriptsize{0.3 : 1.9} & \scriptsize{0.2 : 3.0} & \scriptsize{0.5 : 2.7} &  \scriptsize{1.1 : 1.1}  & \cellcolor[gray]{0.4} & \scriptsize{0.5 : 1.2}  & 	\scriptsize{2.0 : 0.7}  & \scriptsize{1.4 : 0.0}  & \scriptsize{1.2 : 0.4}  & \scriptsize{7 : 11} & \scriptsize{11} & \scriptsize{5} \\ \hline
\scriptsize{Oxsy}  & \scriptsize{0.8 : 0.7} & \scriptsize{0.5 : 1.2} & \scriptsize{1.3 : 3.5} & \scriptsize{1.8 : 0.2}  & \scriptsize{1.2 : 0.5} & \cellcolor[gray]{0.4} & \scriptsize{3.5 : 0.5}  & 	\scriptsize{4.4 : 0.0}  & 	\scriptsize{3.0 : 0.1}  &  \scriptsize{17 : 8}  & \scriptsize{15} & \scriptsize{4} \\ \hline
\scriptsize{Shiraz} & \scriptsize{0.4 : 3.8} & \scriptsize{0.3 : 4.3} & \scriptsize{0.8 : 4.0} & \scriptsize{0.5 : 0.6} & \scriptsize{0.7 : 2.0} & \scriptsize{0.5 : 3.5} & \cellcolor[gray]{0.4} & \scriptsize{0.5 : 0.1} & \scriptsize{0.8 : 1.0} & \scriptsize{6 : 20}  & \scriptsize{5} & \scriptsize{6}\\ \hline
\scriptsize{MT2016} & \scriptsize{0.0 : 5.0} & \scriptsize{0.0 : 3.6} & \scriptsize{0.0 : 5.9} & \scriptsize{0.0 : 0.4} & \scriptsize{0.0 : 1.4}  & \scriptsize{0.0 : 4.4}  & \scriptsize{0.1 : 0.5}  & \cellcolor[gray]{0.4} & \scriptsize{0.0 : 0.0}   &   \scriptsize{0 : 21}  & \scriptsize{2} & \scriptsize{9} \\ \hline
 \scriptsize{FURY}  & \scriptsize{0.2 : 2.5} &  \scriptsize{0.0 : 2.5} & \scriptsize{0.4 : 4.8} & \scriptsize{0.5 : 0.6} & \scriptsize{0.4 : 1.2} & \scriptsize{0.1 : 3.0} & \scriptsize{1.0 : 0.8}  & \scriptsize{0.0 : 0.0} & \cellcolor[gray]{0.4}  & \scriptsize{2 : 17} & \scriptsize{3} & \scriptsize{8} \\ \hline
\end{tabular}
\end{center}
\caption{Evaluation round-robin results (average goals scored and points allocated with discrete scheme), combined for the top 8 teams from RoboCup 2016 and the RoboCup-2015 champion (WE2015). The resultant evaluation ranking is marked with $\mathbf{r^{e}}$.}
\label{te1}
\end{table}

\vspace*{-6mm}
\section{Proposed challenges}

\subsection{Champions Simulation League}

In order to trace the progress of the League over time it is interesting to compare performance of several previous champions, directly competing against each other in a round-robin tournament.   
For example, we evaluated relative performance of six champions of RoboCup-2011 to RoboCup-2016: WrightEagle (WE2011 \cite{we11,Bai2011}, WE2013 \cite{we13,zhang2013decision}, WE2014 \cite{we14}, WE2015 \cite{we15}), HELIOS2012 \cite{helios12} and Gliders2016 \cite{gliders2016tdp,cyb98gliders16}.

\begin{table}[t]
\begin{center}
\begin{tabular}{|c|c|c|c|c|c|c|c|c|c|c|}
 \hline
 & \scriptsize{Gliders2016} &  \scriptsize{ \ WE2015 \ } & \scriptsize{ \ WE2014 \ } & \scriptsize{ \ WE2013 \ } & \scriptsize{HELIOS2012} & \scriptsize{ \ WE2011 \ } &  \scriptsize{Goals} & \scriptsize{Points} & \scriptsize{$\mathbf{r^{l}}$} \\ \hline
\scriptsize{Gliders2016} & \cellcolor[gray]{0.4} & \scriptsize{1.8 : 1.4} & \scriptsize{1.8 : 1.3}  & \scriptsize{1.7 : 0.9}  & \scriptsize{1.2 : 0.1}  & \scriptsize{2.0 : 1.0} &  \scriptsize{9 : 4} & \scriptsize{15} & \scriptsize{1} \\ \hline
\scriptsize{WE2015} & \scriptsize{1.4 : 1.8}  & \cellcolor[gray]{0.4} & \scriptsize{2.5 : 2.5} & \scriptsize{3.0 : 2.5}  & \scriptsize{2.2 : 0.9}  & \scriptsize{4.0 : 2.9}  & \scriptsize{13 : 12} & \scriptsize{8} & \scriptsize{2}\\ \hline
\scriptsize{WE2014} & \scriptsize{1.3 : 1.8}  & \scriptsize{2.5 : 2.5} & \cellcolor[gray]{0.4}  & \scriptsize{2.8 : 2.6}  & \scriptsize{2.3 : 0.8}  & \scriptsize{3.9 : 3.0}     & \scriptsize{13 : 12} & \scriptsize{8} & \scriptsize{3}\\ \hline
\scriptsize{WE2013} & \scriptsize{0.9 : 1.7} & \scriptsize{2.5 : 3.0} & \scriptsize{2.6 : 2.8} & \cellcolor[gray]{0.4} &  \scriptsize{1.9 : 0.9}  &  \scriptsize{2.9 : 3.2}  & \scriptsize{12 : 12}  & \scriptsize{6} & \scriptsize{4}\\ \hline
\scriptsize{HELIOS2012} & \scriptsize{0.1 : 1.2} & \scriptsize{0.9 : 2.2} & \scriptsize{0.8 : 2.3} &  \scriptsize{0.9 : 1.9}  & \cellcolor[gray]{0.4} & \scriptsize{2.6 : 1.8}   & \scriptsize{6 : 9} & \scriptsize{3} & \scriptsize{5} \\ \hline
\scriptsize{WE2011}  & \scriptsize{1.0 : 2.0} & \scriptsize{2.9 : 4.0} & \scriptsize{3.0 : 3.9} & \scriptsize{3.2 : 2.9}  & \scriptsize{1.8 : 2.6} & \cellcolor[gray]{0.4}   &  \scriptsize{12 : 16}  & \scriptsize{1} & \scriptsize{6} \\ \hline
\end{tabular}
\end{center}
\caption{Champions Simulation League round-robin results (average goals scored and points allocated with discrete scheme), for six champions of RoboCup 2011 to 2016. To distinguish WE2015 and WE2014 results, non-rounded scores were used as a tie-breaker. The resultant league ranking with discrete point allocation scheme is marked with $\mathbf{r^{l}}$.}
\label{tel}
\vspace*{-6mm}
\end{table}

The round-robin results over 1000 games, presented in Table \ref{tel}, confirmed the progress of the League over the last six years, with the resultant ranking $\mathbf{r^{l}}$ completely concurring with the chronological ranking $\mathbf{r^{t}}$, i.e.,  $d_1(\mathbf{r^{l}}, \mathbf{r^{t}}) = 0$.

\vspace*{-2mm}
\subsection{Global Challenge}

Another proposal suggests to pit together the best teams from each of the top 6 or 8 participating countries (for example, in 2016 it would have been Australia, Brazil, China, Egypt, Germany, Iran, Japan, Romania), with two ``home-and-away'' games between opponents. There can be 14 games for a home-and-away single-elimination round with 8 teams; or 30 games for a home-and-away double round-robin with 6 teams.  The ``Global Challenge'' will be distinguished from the main competition by playing the games with different parameters, for example, higher noise, or even with random player(s) disconnecting.  In other words, the Global Challenge will focus on resilience of the teams in the face of unexpected conditions.

In each game, the home side would choose a hidden parameter to vary, in order to represent some features of their country (like high altitude in Bolivia or long-distance travel to Australia). These parameters will not be known to the opposition, but would be the same for both teams in that game.   

The full list of possible hidden server parameters may include a significant number (currently, the number of server parameters is 27) and the set of changeable parameters will be agreed in advance. The global challenge mode will be selected via a new parameter, for example, server::global\_challenge\_mode, introduced in the simulation server (server.conf). When the global\_challenge\_mode parameter is set to true, the server will permit  the left side coach (the home side) to send a command like this: (change\_player\_param (param\_1 value) (param\_2 value)$\ldots$ (param\_N value)). 

For example, if the home side chooses to simulate some bad weather conditions or a soggy pitch, these server parameters can be changed: ball\_accel\_max, ball\_decay, \ ball\_rand,\  ball\_speed\_max, \ catch\_probability, \ inertia\_moment, \ kick\_rand, \ player\_rand.

Exploiting their own strong points, and possibly trying to exploit some weak points of the opponent, the home side could change some of the available parameters in a way that creates an advantage. While the adjusted environment will be applied equally to the both teams,  the task of the left side coach (the home team) will be to optimise the choice of the adjusted parameters to maximise the home side advantage. 

\vspace*{-2mm}
\section{Conclusion}

We summarised the results of RoboCup-2016 competition in the  2D Soccer Simulation League, including the main competition and the evaluation round.
The evaluation round confirmed the strength of RoboCup-2015 champion (WrightEagle, i.e. WE2015) in the League, with only eventual finalists of 2016 (Gliders2016 and HELIOS2016) capable of winning against WE2015.  After the RoboCup-2016, we extended this evaluation, over 1000 games for each pair, in a multi-game round-robin tournament which included the top 8 teams of 2016, as well as WE2015. The round-robin results  confirmed that WE2015 would take third place, behind the champion team (Gliders2016) and the runner-up (HELIOS2016). This establishes WE2015 as a stable benchmark for the 2D Simulation League.  In doing so we offered a critique of a particular ranking method (the \emph{continuous} scheme), arguing that the \emph{discrete} scheme is more appropriate.

We then followed with proposing two options to develop the evaluation challenge further.  
The first such possibility  introduces ``The Champions Simulation League'', comprising several previous champions, directly competing against each other in a round-robin tournament. ``The Champions Simulation League'' can systematically trace the advancements in the League, measuring the progress of each new champion over its predecessors. We evaluated The Champions Simulation League with the champions from 2011 to 2016, producing a ranking which completely concurs with the chronological order, and confirming a steady progress in the League. 
Arguably, simulation leagues are the only ones in RoboCup where such an evaluation is possible, given the obvious constraints and difficulties with running such a tournament in robotic leagues.  

Tracing such advances is especially important because different champion teams usually employ different approaches, often achieving a high degree of specialisation in a sub-field of AI, for example, automated hierarchical planning developed by WrightEagle \cite{Bai2011,we13,we14,we15,Bai2015-acm}, opponent modelling studied by HELIOS \cite{helios12}, and human-based evolutionary computation adopted by Gliders \cite{gliders2016tdp,cyb98gliders16}.  Many more research areas are likely to contribute towards improving the League, and several general research directions are recognised as particularly promising: nature-inspired collective intelligence \cite{sayama,Neumann,Hamann}, embodied intelligence \cite{pf06,pol07a,der14,Zahedi}, information theory of distributed cognitive systems \cite{Ay08,Tishby2011,cliff2013towards,liz-inc,frobt.2016.00071,alife}, guided self-organisation \cite{pro09b,DerMartius11,inception}, and deep learning \cite{Bengio,Schmidhuber,Greenwald}.

The other proposed evaluation challenge (``The Global Challenge'')  aims to model environmental conditions during the  games by simulating specific features of different participating countries, such as climate, infrastructure, travel distance, etc.  This, arguably, may increase the realism of the simulated competition, making another small step toward the ultimate Millennium challenge.

\section{Acknowledgments}
A majority of RoboCup 2D Soccer Simulation teams, including the 2016 champion team, Gliders2016, are based on the well-developed code base \emph{agent2d} \cite{agent2d}, release of which has greatly benefited the RoboCup 2D Simulation community. Several teams, including WrightEagle and Oxsy, are independent of \emph{agent2d}.

\bibliographystyle{splncs}

\end{document}